\documentclass[letterpaper,preprintnumbers,prd,twocolumn,nofootinbib,nobibnotes,showpacs]{revtex4}
\usepackage{graphics,epsfig,subfigure}
\usepackage{color}
\usepackage{amsfonts}
\usepackage{mathrsfs}
\usepackage{epsfig}
\usepackage{graphicx}%
\usepackage{dcolumn}
\usepackage{amsmath}
\usepackage{color}
\usepackage{color}

\begin{document}
\renewcommand{\baselinestretch}{1.3}
\newcommand\beq{\begin{equation}}
\newcommand\eeq{\end{equation}}
\newcommand\beqn{\begin{eqnarray}}
\newcommand\eeqn{\end{eqnarray}}
\newcommand\nn{\nonumber}
\newcommand\fc{\frac}
\newcommand\lt{\left}
\newcommand\rt{\right}
\newcommand\pt{\partial}

\allowdisplaybreaks

\title{On black holes with scalar hairs}
\author{Changjun Gao\footnote{gaocj@nao.cas.cn}}\author{Jianhui Qiu\footnote{jhqiu@nao.cas.cn}}

\affiliation{National Astronomical Observatories, Chinese Academy of Sciences, 20A Datun Road, Beijing 100101, China}

\affiliation{School of Astronomy and Space Sciences, University of Chinese Academy of Sciences,
19A Yuquan Road, Beijing 100049, China}

\begin{abstract}
By using the Taylor series method and the solution-generating method, we construct exact black hole solutions with minimally coupled scalar field. We find that the black hole solutions can have many hairs except for the physical mass. These hairs come from the scalar potential. Unlike the mass, there is no symmetry corresponding to these
hairs, thus they are not conserved and one cannot understand them as Noether charges. They arise as coupling constants. Although there are many hairs, the black hole has only one horizon.  The scalar potential becomes negative for sufficient large $\phi$ (or in the vicinity of black hole singularity). Therefore, the no-scalar-hair theorem does not apply to our solutions since the latter does not obey the dominant energy condition. Although the scalar potential becomes negative for sufficient large $\phi$, the black holes are stable to both odd parity perturbations and scalar perturbations. As for even parity perturbations, we find there remains parameter space for the stability of the black holes. Finally, the black hole thermodynamics are developed.
\end{abstract}






\maketitle


\section{Introduction}

The black hole no-hair conjecture \cite{ruf:1971} claims that the final product of gravitational
collapse with any types of matters is the Kerr-Newman black hole which is solely
determined by mass, charge and angular momentum. Except for these three parameters, there exists no other physical quantities, named after the metaphor of ``hair'', for black holes. Since the no-hair conjecture focuses on the fate of gravitational
collapse and not the existence of black hole solutions with
some types of matters, many stationary black hole solutions \cite{biz:1994,bek:1996,vol:1999} with either new
global charges or new non-trivial fields have been found over the past years. It is thus found that there are two types of hairs for black holes.
They are the primary type and the second type. The primary type of hairs include the mass, charge and angular momentum etc.
This type of hairs are subject to the Gauss law and therefore they are global charges. However, the second type of hairs are not
associated to the Gauss law and there is no correspondence with global charges. The scalar hairs of black holes fall into the second type.

Bekenstein has ever formulated the no-scalar-hair theorem \cite{bek:1972l,bek:1972da,bek:1972db} for black holes.
The theorem states that the static black holes with the energy momentum tensor contributed by canonical scalar field and scalar potential
can not have the scalar hairs. According to the classification of nice review paper \cite{car:2015}, three assumptions are made in the theorem.

The first assumption is that the theorem is limited to a canonical kinetic and minimally coupled scalar field with a scalar potential. It is very obvious the first assumption can be reasonably dropped in many scalar-tensor theories, for example, the Brans-Dicke theory \cite{bra:1961}, the Horndeski theory \cite{horn:1974}, the Galileon theory \cite{nic:2009} and the generalized
Galileon theory \cite{def:2009} etc. Therefore, scalar hairy black holes do exist in these theories, for example, in the Yang-Mills gauge theory \cite{vol:1989,biz:1990,gre:1993,mae:1994}, the Skyrme model \cite{luk:1986,bro:1991} and the conformal coupling gravity \cite{bek:1974,bek:1975}.
People have found a lot of hairy black hole solutions by violating the first assumption (see Ref.~\cite{bak:2021} and references therein).

The second assumption is that the scalar field inherits the symmetries of spacetime. Namely, the scalar field is uniquely the function of radial coordinate $r$ in the background of spacetime for static black holes. In fact, this assumption can also be dropped. For example, the scalar field can have a harmonic time dependence if the scalar field is complex. The resulting energy momentum tensor remains time-independent although the scalar field is time dependent. The explicit examples
are the boson stars \cite{sch:2003, lie:2012}. These are self-gravitating, solitonic-like, scalar field configurations which
firstly discussed by Kaup \cite{kan:1968} and Ruffini-Bonazzola \cite{ruf:1969}. Basing on the same mechanism, Herdeiro et al showed that in a series of studies \cite{herd:2014,herd:2015, herd:2016,herd:2019,herd:2021}, the Kerr and Kerr-Newman black holes do have scalar hairs. Furthermore, the scalar hairs measure exactly the conserved and continuous Noether charges. On the other hand, Babichev and Charmousis  et al find that, the scalar hairy black holes could be developed in the Horndeski theory if one allows the scalar field to have a linear time dependence (thus violating the second assumption) \cite{bab:2014,kob:2014,cha:2014}.

The third assumption is that the scalar potential is everywhere non-negative. This assumption is also un-necessary in some scenarios, for example in the
well-known Higgs potential. Actually, abandoning the third assumption is the simplest way to circumvent the no-scalar-hair theorem. Then it is found allowing for the scalar potential to be negative in some range of scalar field can lead to many scalar hairy black holes in the closed form \cite{den:1998,bec:1995,ana:2013,ana:2012,bro:2002,zlo:2005} and numerical form \cite{gub:2005,kle:2013,nuc:2003,cor:2006}. This paper is a continuation of research in this field.

Concretely, by using the Taylor series method and the solution-generating method \cite{cad:2011,cad:2015,wen:2015,cad:2018}, we construct exact black hole solutions with minimally coupled scalar field. The scalar potential is assumed to be infinite series such that it can cover nearly all the known ones. We find that the black hole solutions can have many hairs except for the physical mass. These hairs come from the scalar potential. Unlike the mass, there is no symmetry corresponding to these
hairs, thus they are not conserved and cannot be understood as Noether charges. They arise as merely coupling constants. The solutions require that the black holes have only one horizon although they can possess many hairs. The scalar potential becomes negative for sufficient large $\phi$ (or in the vicinity of black hole singularity). Therefore, the no-scalar-hair theorem does not apply to our solutions since the latter does not obey the dominant energy condition. Although the scalar potential becomes negative for sufficient large $\phi$, the black holes are stable to both odd parity perturbations and scalar perturbations. As for even parity perturbations, we find there remains parameter space for the stability of the black holes. Finally, the black hole thermodynamics is developed. We find there are two pairs of conjugated variables in the first law of thermodynamics. One pair includes the Bekenstein-Hawking entropy and the Hawking temperature. The variables in the other pair are related to the black hole scalar hair and a dimensionless quantity.

The paper is organized as follows. In Section II, we seek for the analytical and asymptotically flat black hole solutions with scalar hairs by using the Taylor series method. Concretely, we expand both the metric functions and the scalar field into infinite series of $1/r$. But in the end of this section, we get an exact solution with finite series. The solution describes a black hole with one scalar hair except for the physical mass. In Section III, by using the solution-generating method, we achieve exact, asymptotically flat,  hairy black hole solutions. The solutions allow black holes having many scalar hairs. The resulting one-scalar-hair black hole solution is exactly the one derived with the Taylor series method. This shows the two methods are consistent with each other. As a consequence of violation of third assumption, the scalar potential is always negative for sufficient large $\phi$ or sufficient small $r$. But this does not mean the solution is not physical. In Section IV and Section V, we explore the metric perturbations and  scalar perturbations to the one-scalar-hair black hole. We find the black holes is always stable to scalar perturbations. As for the metric perturbations, we find it is always stable to odd parity perturbations. For even parity perturbations, there remains parameter space for the stability. Therefore, the black hole solution is physically meaningful. In Section VI, we make an investigation on black hole thermodynamics. We find the scalar hair should enter the first law of thermodynamics as one of conjugated thermodynamic variables. Finally, Section VII gives the
conclusion and discussion. Throughout this paper, we adopt the
system of units in which $G=c=\hbar=1$ and the metric signature
$(-, +, +, +)$.

\section{Taylor series method}
In this section we derive the exact black hole solutions in the presence of canonical scalar field by using the Taylor series method. To this end,
we consider the Einstein theory with non-minimally coupled scalar field which has the action as follows
\begin{equation}
S=\int d^4x\sqrt{-g}\left(R-\frac{1}{2}\nabla_{\mu}\phi\nabla^{\mu}\phi-\sum_{i=1}^{\infty}a_i\phi^{i}\right)\;,
\end{equation}
where the scalar potential $V(\phi)$ is an infinite series. $a_i$ are constants. The physical motivation for taking this form for the
scalar potential is that it can cover nearly all the known scalar potentials.

Variation of the action with respect metric gives the Einstein equations

\begin{equation}
G_{\mu\nu}=-\nabla_{\mu}\phi\nabla_{\nu}\phi+g_{\mu\nu}\left(\frac{1}{2}\nabla_{\alpha}\phi\nabla^{\alpha}\phi+\sum_{i=1}^{\infty}a_i\phi^{i}\right)\;.
\end{equation}
On the other hand, variation of action with respect to the scalar field gives the equation of motion for the scalar field
\begin{equation}
\nabla_{\mu}\nabla^{\mu}\phi-\sum_{i=1}^{\infty}ia_i\phi^{i-1}=0\;.
\end{equation}

We shall look for the static, spherically symmetric black hole solutions in the theory. To this end, we take the ansatz of the metric
\begin{equation}\label{metric}
ds^2=-Udt^2+\frac{\sigma^2}{U}dr^2+r^2d\Omega_2^2\;,
\end{equation}
where $U$ and $\sigma$ are the functions of $r$.
Then the Einstein and scalar field equations give

\begin{eqnarray}
&&4\sigma^{'}+\sigma r\phi^{'2}=0\;,\label{5} \\
&&2r\sigma U^{''}+4\sigma U^{'}-4U\sigma^{'}-2rU^{'}\sigma^{'}-rU\sigma\phi^{'2}\nonumber\\&&-2r\sigma^3\sum_{i=1}^{\infty}a_i\phi^{i}=0\;,\label{6}\\
&&4\sigma^2-4rU^{'}-4U-r^2U\phi^{'2}+2r^2\sigma^2\sum_{i=1}^{\infty}a_i\phi^{i}=0\;,\label{7}\\
&&rU\sigma\phi^{''}+r\sigma U^{'}\phi^{'}-rU\sigma^{'}\phi^{'}+2\sigma U\phi^{'}\nonumber\\&&-r\sigma^3\sum_{i=1}^{\infty}i a_i\phi^{i-1}=0\label{8}\;.
\end{eqnarray}
The prime denotes the derivative with respect to $r$.

We will look for the solutions in the form of series. So we expand $U$, $\phi$ and $U/\sigma^2$ as follows
\begin{eqnarray}\label{12aa}
U&=&1+\sum_{i=1}^{\infty}b_i r^{-i}\;,\ \ \ \ \ \ \ \phi=\sum_{i=1}^{\infty}s_i r^{-i}\;,\nonumber\\
\sigma&=&1+\sum_{i=1}^{\infty}q_i r^{-i}\;,
\end{eqnarray}
such that $\phi$ is asymptotically vanishing and the spacetime is asymptotically Minkowski. Here $b_i$, $s_i$ and $q_i$ are constants to be determined. Substituting Eqs.~(\ref{12aa}) into Eqs.~(\ref{5}-\ref{8}), we find the equations of motion require
\begin{eqnarray}\label{10a}
a_1=a_2=a_3=a_4=0\;,
\end{eqnarray}
and give the solution as follows
\begin{eqnarray}\label{11}
&&U=1+\frac{b_1}{r}\nonumber\\&&+\mathfrak{D_3}\left(s_1,b_1,a_5\right)\cdot\frac{s_1^2}{r^3}\nonumber\\&&
+\mathfrak{D_4}\left(s_1,b_1,a_5,a_6\right)\cdot\frac{s_1^2}{r^4}\nonumber\\&&
+\mathfrak{D_5}\left(s_1,b_1,a_5,a_6,a_7\right)\cdot\frac{s_1^2}{r^5}\nonumber\\&&
\nonumber\\&&+\cdot\cdot\cdot\;,
\end{eqnarray}
with
\begin{eqnarray}\label{11a}
\mathfrak{D_3}&=&\frac{1}{24}\cdot\left(b_1+4a_5s_1^3\right)\;,\\
\mathfrak{D_4}&=&\frac{1}{24}\cdot\left(2a_6s_1^4+25a_5^2s_1^6-b_1^2\right)\;,\\
\mathfrak{D_5}&=&\frac{1}{1920}\cdot\left(96a_7s_1^5-2200b_1a_5^2s_1^6-320a_5b_1^2s_1^3\right.\nonumber\\&&\left.-144a_6b_1s_1^4
+10000a_5^3s_1^9+72b_1^3+56a_5s_1^5\right.\nonumber\\&&\left.+1920a_5s_1^7a_6
+9b_1s_1^2\right)\;;
\end{eqnarray}
\begin{eqnarray}\label{15}
&&\phi=\frac{s_1}{r}+\mathfrak{E_2}\left(s_1,b_1,a_5\right)\cdot\frac{s_1}{r^2}\nonumber\\&&+\mathfrak{E_3}\left(s_1,b_1,a_5,a_6\right)
\cdot\frac{s_1}{r^3}\nonumber\\&&+\mathfrak{E_4}\left(s_1,b_1,a_5,a_6,a_7\right)\cdot\frac{s_1}{r^4}\nonumber\\&&+\cdot\cdot\cdot\;,
\end{eqnarray}
with
\begin{eqnarray}
\mathfrak{E_2}&=&\frac{1}{2}\cdot\left(5a_5s_1^3-b_1\right)\;,\\
\mathfrak{E_3}&=&\frac{1}{24}\cdot\left(-80a_5s_1^3b_1
+200a_5^2s_1^6+s_1^2\right.\nonumber\\&&\left.+8b_1^2+24a_6s_1^4\right)\;,\\
\mathfrak{E_4}&=&\frac{1}{72}\cdot\left(-1300a_5^2s_1^6b_1+2125a_5^3s_1^9
+32s_1^5a_5\right.\nonumber\\&&\left.+265a_5s_1^3b_1^2+570a_5s_1^7a_6+
42a_7s_1^5\right.\nonumber\\&&\left.-6s_1^2b_1
-18b_1^3-144b_1s_1^4a_6\right)\;;
\end{eqnarray}

\begin{eqnarray}\label{19}
&&\sigma=1+\frac{s_1^2}{8r^2}\nonumber\\&&+\mathfrak{F_3}\left(s_1,b_1,a_5\right)\cdot\frac{s_1^2}{r^3}\nonumber\\&&
+\mathfrak{F_4}\left(s_1,b_1,a_5,a_6\right)\cdot\frac{s_1^2}{r^4}\nonumber\\&&
+\mathfrak{F_5}\left(s_1,b_1,a_5,a_6,a_7\right)\cdot\frac{s_1^2}{r^5}\nonumber\\&&
\nonumber\\&&+\cdot\cdot\cdot\;,
\end{eqnarray}
with

\begin{eqnarray}\label{13a}
\mathfrak{F_3}&=&\frac{1}{6}\cdot\left(-b_1+5a_5s_1^3\right)\;,\\
\mathfrak{F_4}&=&\frac{3}{128}\cdot\left(s_1^2+16a_6s_1^4-80a_5s_1^3b_1
+200a_5^2s_1^6\right.\nonumber\\&&\left.+8b_1^2\right)\;,\\
\mathfrak{F_5}&=&\frac{1}{180}\cdot\left(42a_7s_1^5-2650b_1a_5^2s_1^6
+535a_5b_1^2s_1^3\right.\nonumber\\&&\left.
+840a_5s_1^7a_6-198a_6b_1s_1^4
+4375a_5^3s_1^9\right.\nonumber\\&&\left.-36b_1^3+62a_5s_1^5-12b_1s_1^2\right)\;.
\end{eqnarray}
Here $\mathfrak{D_i},\mathfrak{E_i},\mathfrak{F_i}$ represent the functions of corresponding parameters. We do not give the expressions of $\mathfrak{D_{6,7,8,\cdot\cdot\cdot}},\mathfrak{E_{5,6,7,\cdot\cdot\cdot}},\mathfrak{F_{6,7,8,\cdot\cdot\cdot}}$  because they are rather lengthy. Observing this solution, we find there are only two integration constants, $b_1$ and $s_1$. $s_1$ can be understood as the scalar charge. Let $b_1=-2M$. Then $M$ is exactly the mass of the black hole. When the scalar charge $s_1$ vanishes, the solution restores to the Schwarzschild solution. In all, the black hole solution have two hairs, the mass $M$ and the scalar charge $s_1$. In contrast, $a_i$ are understood as the coupling constants. Since $a_i$ appear in the metric functions and they could play great role in the structure of black hole spacetime, it is reasonable to regard them as the hairs of black holes. Therefore, a black hole can have many hairs.

Eqs.~(\ref{10a}) tells us that if the scalar potential $V(\phi)$ is chosen as
\begin{eqnarray}
V=a_1\phi+a_2\phi^2+a_3\phi^3+a_4\phi^4\;,
\end{eqnarray}
the resulting black hole solution is nothing but the Schwarzschild solution.

The expressions of $U$, $\phi$ and $\sigma $ (Eqs.(\ref{11},\ref{15},\ref{19})) are infinite series. In principle, they can cover a vast of the solutions, for example, \cite{den:1998,ana:2014,gua:2014,tar:2014}. In order to get finite series, for example, with only two terms present in $\sigma$, we can set
\begin{eqnarray}
&&\mathfrak{F_3}=0\ \ \Rightarrow \ \ a_5=\mathfrak{H_5}\left(s_1,b_1\right)\;,\nonumber\\
&&\mathfrak{F_4}=0\ \ \Rightarrow \ \ a_6=\mathfrak{H_6}\left(s_1,b_1,a_5\right)\;,\nonumber\\
&&\mathfrak{F_5}=0\ \ \Rightarrow \ \ a_7=\mathfrak{H_7}\left(s_1,b_1,a_5,a_6\right)\;,\nonumber\\
&&\cdot\cdot\cdot\;,
\end{eqnarray}
where $\mathfrak{H_i}$ are the functions of corresponding parameters.
Then we are left with only two free parameters, namely the mass $M$ and the scalar charge $s_1$. Then the resulting black hole solution is
\begin{eqnarray}
&&{\sigma}=1+\frac{s_1^2}{8r^2}\;,\label{25}\\
&&{U}=1-\frac{2M}{r}-\frac{3Ms_1^2}{20r^3}-\frac{s_1^4}{192r^4}\;,\label{26}\\
&&\phi=\frac{s_1}{r}-\frac{s_1^3}{48r^3}+\cdot\cdot\cdot\;.\label{27}
\end{eqnarray}
Now the expressions of both $\sigma$ and $U$ are finite series while $\phi$ remains infinite. This is the solution we are interested in. In the next section, we shall meet this solution once again by using the solution-generating method.

\section{solution-generating method}
In this section, we explore the exact black hole solutions in the presence of canonical scalar field by using the solution-generating method \cite{cad:2011,cad:2015,wen:2015,cad:2018}. To this end, we consider the action as follows
\begin{equation}
S=\int d^4x\sqrt{-g}\left[R-\frac{1}{2}\nabla_{\mu}\phi\nabla^{\mu}\phi-V\left(\phi\right)\right]\;,
\end{equation}
where the scalar potential $V(\phi)$ is assumed to be not an infinite series, but a scalar function to be determined. The reason for this point is that we will explore the black hole solution with the solution-generating method.

Variation of the action with respect metric gives the Einstein equations
\begin{equation}
G_{\mu\nu}=-\nabla_{\mu}\phi\nabla_{\nu}\phi+g_{\mu\nu}\left[\frac{1}{2}\nabla_{\alpha}\phi\nabla^{\alpha}\phi+V\left(\phi\right)\right]\;.
\end{equation}
On the other hand, variation of action with respect to the scalar field gives the equation of motion for the scalar
\begin{equation}
\nabla_{\mu}\nabla^{\mu}\phi-V_{,\phi}=0\;.
\end{equation}

Our goal is to look for the static, spherically symmetric black hole solutions in the theory. For this purpose, we take the ansatz of the metric same as Eq.~(\ref{metric}).
Then the Einstein and scalar field equations give
\begin{eqnarray}
&&4\sigma^{'}+\sigma r\phi^{'2}=0\;,\label{31} \\
&&2r\sigma U^{''}+4\sigma U^{'}-4U\sigma^{'}-2rU^{'}\sigma^{'}-rU\sigma\phi^{'2}\nonumber\\&&-2r\sigma^3V=0\;,\label{32}\\
&&4\sigma^2-4rU^{'}-4U-r^2U\phi^{'2}+2r^2\sigma^2V=0\;,\label{33}\\
&&rU\sigma\phi^{''}+r\sigma U^{'}\phi^{'}-rU\sigma^{'}\phi^{'}+2\sigma U\phi^{'}\nonumber\\&&-r\sigma^3V_{,\phi}=0\label{34}\;.
\end{eqnarray}
We notice that only three of the four equations of motion are independent because of the Bianchi identities. But we have four variables, $U$, $\sigma$, $\phi$ and $V$.
We usually assign the expression of scalar potential in advance. Then the equations of motion are closed. But in practice, one can assign anyone of the four variables in advance. We find that the equations of motion become considerable simple once we assume the expression of $\sigma$ in advance. To show this point, we combine Eqs.~(\ref{32}-\ref{34}) and find
\begin{eqnarray}
-2rU\sigma^{'}+2U\sigma-r^2\sigma U^{''}+r^2U^{'}\sigma^{'}-2\sigma^3=0\;.
\end{eqnarray}
Solving this equation, we obtain
\begin{eqnarray}\label{38}
U=r^2\left(c_1+\int\frac{c_2\sigma}{r^4}dr-\int\frac{2\sigma\int{\sigma dr}}{r^4}dr\right)\;.
\end{eqnarray}
Here $c_1$ and $c_2$ are integration constants. It is easy to complete the integrations if we assume $\sigma$ has the expression of power-series
\begin{eqnarray}
\sigma=1+\sum_{i=1}^{\infty}w_i r^{-i}\;,
\end{eqnarray}
where $w_i$ are constants. We emphasize that we are interested in the asymptotically flat spacetime solution. Therefore we require $c_1=0$.

\begin{figure}[htbp]
	\centering
	\includegraphics[width=8cm,height=6cm]{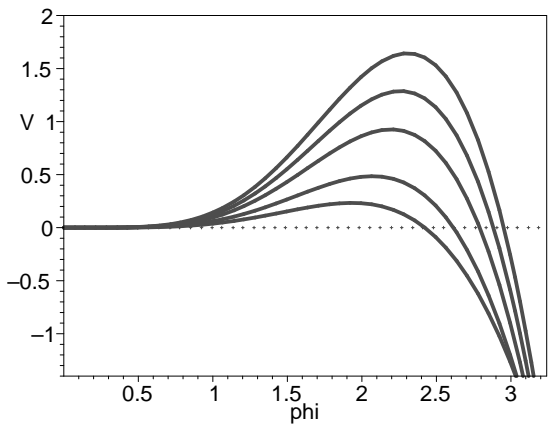}
	\caption{The scalar potential $V(\phi)$ for $M=1$ and $w_2=0.18,\ 0.20, \ 0.23, \ 0.3, \ 0.4$, respectively, from top to bottom.}
	\label{Vphi.eps}
\end{figure}

In the first place, we consider
\begin{eqnarray}
\sigma=1+\frac{w_1}{r}+\frac{w_2}{r^2}\;.
\end{eqnarray}
Then we obtain from Eq.~(\ref{38})
\begin{eqnarray}
U&=&1+\frac{1}{r}\cdot\left(20c_2+\frac{8}{9}w_1+\frac{2}{3}w_1\ln{r}\right)\nonumber\\&&+\frac{1}{r^2}\cdot\left(15c_2w_1+\frac{1}{8}w_1^2+\frac{1}{2}w_1^2\ln{r}
\right)\nonumber\\&&+\frac{1}{r^3}\cdot\left(12c_2w_2-\frac{8}{25}w_1w_2+\frac{2}{5}w_1\ln{r}\right)\nonumber\\&&+\frac{1}{r^4}
\cdot\left(-\frac{1}{3}w_2^2
\right)\;.
\end{eqnarray}
For sufficient large $r$, we should get the Schwarzschild solution. Thus we should let
\begin{eqnarray}
w_1=0\;,\ \ \ c_2=-\frac{M}{10}\;,
\end{eqnarray}
and eventually we obtain
\begin{eqnarray}
U&=&1-\frac{2M}{r}-\frac{{6}Mw_2}{5r^3}-\frac{{w_2^2}}{3r^4}\;.\label{41}\\
\sigma&=&1+\frac{w_2}{r^2}\;,\label{42}\\
\phi&=&2\sqrt{2}\ \textrm{arctanh}\frac{\sqrt{w_2}}{\sqrt{{r^2+w_2}}}\;,\label{43}\\
V&=&\frac{4}{15}\cdot\frac{\sinh^5{\frac{\phi}{2\sqrt{2}}}}{w_2^2\cosh^6{\frac{\phi}{2\sqrt{2}}}}
\left[-5w_2\sinh^3\frac{\phi}{2\sqrt{2}}\right.\nonumber\\&&\left.-15w_2\sinh\frac{\phi}{2\sqrt{2}}+12M\sqrt{w_2}\right]\;.
\end{eqnarray}
Expanding Eq.~(\ref{43}) into Taylor series of $1/r$ and comparing Eqs.~(\ref{41},\ref{42},\ref{43}) with Eqs.~(\ref{25},\ref{26},\ref{27}), we find they are exactly identical provided that $w_2=s_{1}^2/8$. So they are nothing but the solution we present in section II.
Observing the expression of $\phi$, we find we should require
\begin{eqnarray}
w_2>0\;.
\end{eqnarray}
Thus the scalar field $\phi$ is always nonnegative. The linear stability of the asymptotic Schwarzschild solution is guaranteed when one takes into account of the fact
\begin{eqnarray}
V_{,\phi\phi}|_{\phi=0}=0\;.
\end{eqnarray}
The equation of horizon is
\begin{eqnarray}
U=0\;.
\end{eqnarray}
It gives a fourth order equation but it  has only one positive root. So this black hole has only one event horizon. The physical singularity locates at $r=0$. In Fig.~(\ref{Vphi.eps}) we plot the scalar potential $V(\phi)$. It is apparent the potential is negative for large $\phi$ (or in the vicinity of black hole singularity). Therefore, the no-hair theorem \cite{heu:1996,may:1996} simply does not apply to our solution since the latter does not  obey the dominant energy condition.

\begin{figure}[htbp]
	\centering
	\includegraphics[width=8cm,height=6cm]{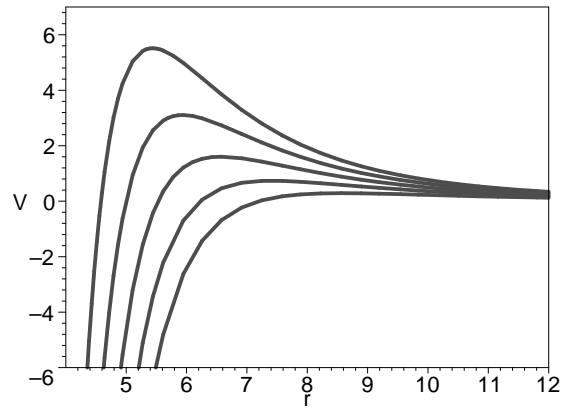}
	\caption{The scalar potential $V(r)$ for $M=1, \ w_3=w_4=w_5=1$ and $w_2=0.01,\ 0.05, \ 0.09, \ 0.13, \ 0.17$, respectively, from top to bottom.}
	\label{Vr.eps}
\end{figure}

Secondly, we can derive the other black hole solutions. For example, if we take
\begin{eqnarray}
\sigma=1+\frac{w_2}{r^2}+\frac{w_3}{r^3}+\frac{w_4}{r^4}+\frac{w_5}{r^5}\;,\label{47}
\end{eqnarray}
then we obtain
\begin{eqnarray}\label{48}
U&=&1-\frac{2M}{r}+\left(-\frac{6}{5}Mw_2+\frac{1}{5}w_3\right)\cdot\frac{1}{r^3}\nonumber\\&&
+\left(\frac{2}{9}w_4-\frac{1}{3}w_2^2-Mw_3\right)\cdot\frac{1}{r^4}\nonumber\\&&
+\left(-\frac{3}{7}w_2w_3+\frac{3}{14}w_5-\frac{6}{7}Mw_4\right)\cdot\frac{1}{r^5}\nonumber\\&&+\left(-\frac{1}{3}w_2w_4
-\frac{1}{8}w_3^2-\frac{3}{4}Mw_5\right)\cdot\frac{1}{r^6}\nonumber\\&&+\left(-\frac{5}{27}w_3w_4-\frac{5}{18}w_2w_5\right)\cdot\frac{1}{r^7}
\nonumber\\&&+\left(-\frac{3}{20}w_3w_5-\frac{1}{15}w_4^2\right)\cdot\frac{1}{r^8}\nonumber\\&&-\frac{7}{66}w_4w_5\cdot\frac{1}{r^9}-\frac{w_5^2}{24}
\cdot\frac{1}{r^{10}}\;,
\end{eqnarray}
and
\begin{eqnarray}
\phi=\int 2\sqrt{\frac{2w_2r^3+3w_3r^2+4w_4r+5w_5}{\left(r^5+w_2r^3+w_3r^2+w_4r+w_5\right)r^2}}dr \;.\label{49}
\end{eqnarray}
Eqs.~(\ref{47},\ref{48},\ref{49}) constitute a new exact black hole solution. Here we do not give the expressions of $V$  because it is rather complicated. But in Fig.~(\ref{Vr.eps}) we plot the scalar potential $V(r)$. It is apparent the potential becomes negative when one approaches the black hole singularity.
In order that the expression of $\phi$ always makes sense when $r\geq 0$, we should require all the $w_i$ are positive. Then the black hole has only one event horizon.
\section{perturbations}
The problem for dealing with black hole stability to perturbations was firstly solved by Regge and Wheeler \cite{reg:1957} in 1957. They showed Schwarzschild black hole is stable to odd parity perturbations. Then in 1969, Vishveshwara \cite {vis:1970a, vis:1970b} showed that Schwarzschild black hole is also stable to even parity perturbations. Finally, Zerrili examined these conclusions in a unified manner \cite{zer:1970}. In this section, we shall make the investigation on stability problem for black hole which contains an additional scalar field. The procedure for carrying out this task has been developed in \cite{den:1998}. In order to follow the notation of \cite{den:1998}, we should rewrite the metric, Eqs.~(\ref{41},\ref{42})  from the $r$ coordinate to $x$ coordinate in the form
\begin{eqnarray}
ds^2=-U\left(x\right)dt^2+\frac{1}{U\left(x\right)}dx^2+f\left(x\right)^2d\Omega_2^2\;,\label{51}
\end{eqnarray}
with
\begin{eqnarray}
U\left(x\right)&=&1-\frac{2M}{f\left(x\right)}-\frac{6}{5}\cdot\frac{Mw_2}{f\left(x\right)^3}-\frac{1}{3}
\cdot\frac{w_2^2}{f\left(x\right)^4}\;,\\
f\left(x\right)&\equiv&\frac{1}{2}x+\frac{1}{2}\sqrt{x^2+4w_2}\;,
\end{eqnarray}
via coordinate transformation
\begin{eqnarray}
r=\frac{1}{2}x+\frac{1}{2}\sqrt{x^2+4w_2}\;.
\end{eqnarray}
Now $x$ is in the interval of $(-\infty,\ +\infty)$. $x=-\infty$ and $x=+\infty$ represent the black hole singularity $r=0$ and spatial infinity $r=+\infty$, respectively.

The total metric $g^{tot}_{\mu\nu}$ is decomposed into two parts, i.e the background metric $g_{\mu\nu}$ and the perturbations $h_{\mu\nu}$,
\begin{equation}\label{metr-tot}
g^{tot}_{\mu\nu}=g_{\mu\nu}+\delta g_{\mu\nu}\,,\end{equation}
with $|\delta g_{\mu\nu}|\ll|g_{\mu\nu}|$.
On the other hand, the total scalar field $\phi^{tot}$ is decomposed into the background field $\phi$ and the perturbations $\delta\phi$,
\begin{equation}\label{Phi-tot}
\phi^{tot}=\phi+\delta\phi\,,\end{equation}
with $|\delta\phi|\ll|\phi|$. The perturbations are regraded as independent fields, evolving in the background of the static spherically symmetric hairy black hole. This implies their energy, angular momentum and parity are all conserved. Therefore, the modes with different energy, angular momentum or parity do not mix with the evolution of time. So each mode can be dealt with separately. By introducing a factor of $e^{I\omega t}$ in each mode of the decomposition, the energy $\omega$ is fixed. Likewise, the introduction of factor, spherical harmonic function $Y_{l}^{m}$ in each mode fixes the angular momentum. Then we are left with pure radial degrees of freedom and they are divided into two parts with the odd parity and even parity, respectively.

Since the spherical harmonic functions $Y_{l}^{m}$ already have definite parity $(-1)^{l}$, it is not necessary to make parity separation for the scalar perturbations. On the contrary, the metric perturbations possess directional information due to their tensor nature. Therefore, the parity separation must be performed for the tensor perturbations. The resulting equations of motion are simplified by gauging away four radial degrees of freedom. What is more, with a rotation performed on each mode, the spherical harmonic functions $Y_{l}^{m}$ is simplified to be the Legendre polynomial $P_{l}(\cos{\theta})$. This is called the canonical decomposition \cite{reg:1957}.
They are eventually given by
\begin{itemize}
    \item for odd parity:
\begin{eqnarray}
\delta g^{\rm odd}_{\mu \nu}= \left[
\begin{array}{cccc}
0 & 0 &0 & \epsilon h_8(r)
\\ 0 & 0 &0 & \epsilon h_9(r)
\\ 0 & 0 &0 & 0
\\ \epsilon h_8(r) & \epsilon h_9(r) &0 &0
\end{array}\right] e^{I\omega t}
\sin\theta\,\partial_\theta P_l\,,\end{eqnarray}
\begin{eqnarray}\delta\phi=0\,,\end{eqnarray}
\item for even parity:
\begin{eqnarray}\label{h-even}
\delta g^{\rm even}_{\mu \nu}= \left[
 \begin{array}{cccc}
 \epsilon h_1(r)& \epsilon h_2(r) &0 & 0
\\ \epsilon h_2(r) & \epsilon h_3(r)  &0 & 0
\\ 0 & 0 &\epsilon h_7(r) & 0
\\ 0 & 0 &0 & \epsilon h_7(r)\,\sin^2\theta
\end{array}\right] e^{I\omega t}
P_{l}\,,\nonumber\\\end{eqnarray}
\begin{equation}\label{phi-even}
\delta\phi(r)={\epsilon \psi}(r)\,e^{I\omega t}P_{l}\,.\end{equation}
\end{itemize}

Here the constant $\epsilon$ denotes the order of perturbations. The advantage of introduction for $\epsilon$ is that what we should do is just to collect the terms with the order of $\epsilon^2$ in the Lagrangian or the terms with the order of $\epsilon$ in the equations of motion and eventually we let $\epsilon=1$.

\subsection{odd parity}
Substituting now Eqs. (\ref{metr-tot}) and (\ref{Phi-tot}) with the odd parity into the field equations and retaining the terms only linear in $\epsilon$, we obtain  finally two independent equations after setting $\epsilon=1$,
\begin{eqnarray}\label{odd1}
&&\left\{h_9\left[\frac{l\left(l+1\right)-2}{2f^2}-\frac{\omega^2}{2U}\right]+I\omega h_8\frac{f^{'}}{Uf}-\frac{I\omega}{2U}h_8^{'}\right\}\nonumber\\&&
\cdot\sin\theta \partial_{\theta}P_{l}e^{I\omega t}=0\;,
\end{eqnarray}
\begin{eqnarray}\label{odd2}
&&\left[\frac{I\omega}{U}h_8-\left(Uh_9\right)^{'}\right]\nonumber\\&&
\cdot\left[\frac{1}{2}l\left(l+1\right)\sin{\theta}P_l+\cos{\theta}\partial_{\theta}P_l\right]e^{I\omega t}=0\;,
\end{eqnarray}
where the prime denotes the derivative with respect to $x$. We notice that when $l=0$, we have the Legendre polynomial $P_0=1$. Then both Eq.~(\ref{odd1}) and Eq.~(\ref{odd2}) become identities. In other words, there is no monopole perturbations for odd parity. On the other hand, when $l=1$ we have the Legendre polynomial $P_1=\cos\theta$. Then Eq.~(\ref{odd2}) becomes an identity. We are left with unique differential equation Eq.~(\ref{odd1}) but with two degrees of freedom, $h_8$ and $h_9$. Therefore, the question turns out to be incomplete. In fact, with the help of gauge transformation
\begin{eqnarray}
\tilde{x}^{i}=x^{i}+\delta^{i}_{\varphi}\frac{I}{\omega f^2}{e^{I\omega t}}h_8\;,
\end{eqnarray}
and a redefinition of $h_9$, we are able to let $h_8$ totally vanishes. Then the remaining Eq.~(\ref{odd1}) also gives vanishing $h_9$. This shows there is no polar-pole mode for the odd parity, either. Up to this pint, we conclude that the lowest angular momentum of odd parity perturbations is $l=2$ because the angular parts of Eq.(\ref{odd1}) and Eq.~(\ref{odd2}) never vanish when $l\geq 2$. Therefore, we can focus on the radial parts of Eq.(\ref{odd1}) and Eq.~(\ref{odd2}) in the next. Then we obtain
\begin{eqnarray}\label{odd3}
&&\left\{\frac{2UU^{'}f^{'}}{f}-U^{'2}-UU^{''}-\frac{U}{f^2}\left[l\left(l+1\right)-2\right]\right\}h_9\nonumber\\&&
+\left(\frac{2U^2f^{'}}{f}-3UU^{'}\right)h_9^{'}-U^2h_{9}^{''}=\omega^2h_9\;.
\end{eqnarray}
By introducing a new dynamical variable $\Psi$ taking the place of $h_9$ and the tortoise coordinate $r_{\ast}$ taking the place of $x$ as follows
\begin{eqnarray}
\Psi\equiv \frac{Uh_9}{f}\;,\ \ \ \ r_{\ast}\equiv\int\frac{1}{U\left(x\right)}dx\;,
\end{eqnarray}
we find the perturbation equation Eq.~(\ref{odd3}) is simplified to be
\begin{eqnarray}
-\partial^2_{r_{\ast}}\Psi\left(r_{\ast}\right)+V_{eff}\left(r_{\ast}\right)\Psi\left(r_{\ast}\right)=\omega^2\Psi\left(r_{\ast}\right)\;,
\end{eqnarray}
where
\begin{eqnarray}
V_{eff}\left(r_{\ast}\right)=\frac{U}{f^2}\left[l\left(l+1\right)-2\right]+f\left(\partial^2_{r_{\ast}}\frac{1}{f}\right)\;,
\end{eqnarray}
or
\begin{eqnarray}\label{oddpot}
V_{eff}\left(r_{\ast}\right)=\frac{U}{f^2}\left[l\left(l+1\right)-2\right]+fU\partial_{x}\left(U\partial_{x}\frac{1}{f}\right)\;.
\end{eqnarray}
The tortoise coordinate $r_{\ast}$ is a regular, monotonic function of $x$ in the exterior of the black hole because of $U>0$ there. Since $U$ tends to $1$ for large distances, $r_{\ast}$ approaches positive infinity when $x\rightarrow+\infty$. On other hand, in the vicinity of the horizon, $U$ becomes zero such that $r_{\ast}$ tends to negative infinity. Therefore, $r_{\ast}$ covers the whole exterior of the black hole. We note that the lowest angular momentum $l$ for the effective potential is $l=2$ because the corresponding perturbations are vanishing for $l=0,\ 1$.

\begin{figure}[htbp]
	\centering
	\includegraphics[width=8cm,height=6cm]{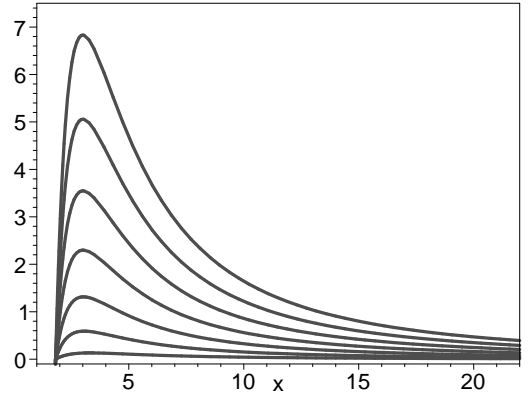}
	\caption{The effective potential $V_{eff,odd}$ of odd parity for $M=1, \ w_2=1$ and $l=2,\ 4, \ 6, \ 8, \ 10,\ 12,\ 14$, respectively, from down to up. It is positive everywhere outside the black hole.}
	\label{Vx.eps}
\end{figure}

Kobayashi et. al \cite{kob:2012} have completed the study of black hole perturbations for odd parity in the Horndeski theory which covers the Einstein-scalar theory studied in this paper. Therefore, to study the stability problem, we may refer their results and notations. Then we find
\begin{eqnarray}
\mathcal{F}=1\;,\ \ \mathcal{G}=1\;,\ \ \mathcal{H}=1\;.
\end{eqnarray}
In order to avoid gradient instability, one requires
\begin{eqnarray}
\mathcal{F}>0\;.
\end{eqnarray}
On the other hand, in order to avoid the presence of ghost, one requires
\begin{eqnarray}
\mathcal{G}>0\;.
\end{eqnarray}
It is apparent the two conditions are all satisfied. The squared propagation speeds of gravitational waves along the radial direction,
$c_r^2$ and the angular direction $c_{\theta}^2$ are found to be
\begin{eqnarray}
c_{r}^2=\frac{\mathcal{G}}{\mathcal{F}}=1\;,\ \ \ c_{\theta}^2=\frac{\mathcal{G}}{\mathcal{H}}=1\;.
\end{eqnarray}
Namely, they are exactly the square of speed for light. The above conditions of  $\mathcal{F}>0,\mathcal{G}>0$ and $\mathcal{H}>0$ are all necessary, but not sufficient for the stability of black holes. The sufficient condition for stability is

\begin{eqnarray}
V_{eff,odd}\left(r_{\ast}\right)\geq 0\;,
\end{eqnarray}
in the exterior of black holes. By using Eq.~(\ref{oddpot}), we find it is indeed the case regardless of the values for $M$ and $w_2$ when $l\geq 2$. One only requires $M>0$ and $w_2>0$.  As an example, we plot the effective potential $V_{eff}$ in terms of $x$ in Fig.~(\ref{Vx.eps}). The potential is positive everywhere and tends to zero at both the infinity and the horizon. Therefore, we conclude that the black hole is stable to odd parity perturbations.

Once the black hole is perturbed it would evolve into three stages by emitting gravitational waves: 1) a relatively short period of initial outburst of radiation,
2) a long period of damping proper oscillations, dominated by the so-called quasinormal
modes, 3) at very large time the quasinormal modes are suppressed by power-law or exponential late-time tails. It is found the dominating contribution
to gravitational waves is the quasinormal mode with the lowest frequency: the fundamental mode. So in the next, we turn to the evaluation of quasinormal frequencies for the odd parity perturbations by using the third-order WKB approximation, a numerical and perhaps the most popular method, devised by Schutz, Will and Iyer \cite{will:1985,will:1987,iyer:1987}. This method has been used extensively in evaluating quasinormal frequencies of various black holes. For an incomplete list see \cite {quasi:99} and references therein.

The quasinormal frequencies are given by
\begin{eqnarray}
\omega^2=V_0+\Lambda\sqrt{-2V_0^{''}}-i\nu\left(1+\Omega\right)\sqrt{-2V_0^{''}}\;,
\end{eqnarray}
where $\Lambda$ and $\Omega$ are

\begin{eqnarray}
\Lambda&=&\frac{1}{\sqrt{-2V_0^{''}}}\left\{\frac{V_0^{(4)}}{V_0^{''}}\left(\frac{1}{32}+\frac{1}{8}\nu^2\right)
\right.\nonumber\\&&\left.-\left(\frac{V_0^{'''}}{V_0^{''}}\right)^2\left(\frac{7}{288}+\frac{5}{24}\nu^2\right)\right\}\;,\\
\Omega&=&\frac{1}{\sqrt{-2V_0^{''}}}\left\{\frac{5}{6912}\left(\frac{V_0^{'''}}{V_0^{''}}\right)^4\left(77+188\nu^2\right)
\right.\nonumber\\&&\left.-\frac{1}{384}\left(\frac{V_0^{'''2}V_0^{(4)}}{V_0^{''3}}\right)\left(51+100\nu^2\right)
\right.\nonumber\\&&\left.+\frac{1}{2304}\left(\frac{V_0^{(4)}}{V_0^{''}}\right)^2\left(67+68\nu^2\right)\right.\nonumber\\&&\left.
+\frac{1}{288}\left(\frac{V_0^{'''}V_0^{(5)}}{V_0^{''2}}\right)\left(19+28\nu^2\right)\right.\nonumber\\&&\left.-\frac{1}{288}\left(\frac{V_0^{(6)}}{V_0^{''}}
\left(5+4\nu^2\right)\right)\right\}\;,
\end{eqnarray}
and
\begin{eqnarray}
\nu=n+\frac{1}{2}\;,\ \ \ \ V_0^{(s)}=\frac{d^sV}{dr_{\ast}^s}|_{r_{\ast}=r_p}\;,
\end{eqnarray}
$n$ is overtone number and $r_p$ corresponds to the peak of the effective
potential. It is pointed that \cite{car:04} that the accuracy of the WKB method depends on the multipolar
number $l$ and the overtone number $n$. The WKB approach is consistent with the numerical method very well provide that  $l>n$.
Therefore we shall present the quasinormal frequencies of scalar perturbation for $n=0$ and $l=2,3,4,5,6,7$, respectively.

The fundamental quasinormal frequencies of the odd parity perturbations are given in table I. From the table we see that with the increasing
of scalar charge $w_2$,the damping time ($\sim1/\textrm{Im}(\omega)$) of gravitational waves becomes longer and longer. On the other hand, the period of oscillation of gravitational waves ($\sim2\pi/\textrm{Re}(\omega)$) also becomes longer and longer.

\begin{table*}[tbp]
\centering
\begin{tabular}{lcccccccc}
\hline
$w_{_2}$ & $\omega\ (l=2)$ & $\omega\ (l=3)$ & $\omega\ (l=4)$ & $\omega\ (l=5)$ & $\omega\ (l=6)$ & $\omega\ (l=7)$ &\\ \hline
0.01 &0.3739-0.0933*I &0.5991-0.0944*I &0.8086-0.0950*I &1.0116-0.0953*I &1.2112-0.0955*I &1.4088-0.0956*I  \\
0.1 &0.3714-0.0926*I &0.5954-0.0937*I &0.8037-0.0944*I &1.0055-0.0947*I &1.2039-0.0949*I &1.4004-0.0951*I  \\
0.3 &0.3661-0.0911*I &0.5875-0.0925*I &0.7933-0.0931*I &0.9926-0.0935*I &1.1885-0.0937*I &1.3825-0.0939*I  \\
0.5 &0.3612-0.0898*I &0.5800-0.0912*I &0.7834-0.0920*I &0.9803-0.0924*I &1.1740-0.0926*I &1.3656-0.0927*I  \\
0.7 &0.3566-0.0885*I &0.5730-0.0901*I &0.7741-0.0909*I &0.9687-0.0913*I &1.1601-0.0915*I &1.3496-0.0916*I  \\
\hline
\end{tabular}
\caption{The quasinormal frequencies of gravitational waves for odd parity.}
\end{table*}

\subsection{even parity}
In this section, we turn to the metric perturbations for even parity. Similar to the case of odd parity,  we substitute Eqs. (\ref{metr-tot}) and (\ref{Phi-tot}) with the even parity into the field equations and keep the terms only linear in $\epsilon$. The resulting equations of motion are rather complicated and we have not been able to achieve a final Schrodinger-like equation \cite{den:1998,kob:2012} except for the case of $l=0$. When $l=0$, one could gauge away two degrees of freedom by putting
\begin{eqnarray}
h_1=0\;,\ \ \ \ h_7=0\;.
\end{eqnarray}
In the framework of this gauge, the $(0,1)$ component of the Einstein equations becomes significantly simple and it can be solved for $h_3$
\begin{eqnarray}
h_3=-\frac{f\phi^{'}}{2Uf^{'}}\psi\;.
\end{eqnarray}
On the other hand, the combination of $(2,2)$ and $(2,3)$ components of Einstein equations gives
\begin{eqnarray}
h_2&=&\frac{I}{16\omega}\left(\frac{4fU\phi^{''}}{f^{'}}+\frac{f^2U\phi^{'3}}{f^{'2}}-4U\phi^{'}\right)\psi\nonumber\\&&+\frac{IUf\phi^{'}}{4\omega f^{'}}\psi^{'}\;.
\end{eqnarray}
By using the expressions of $h_2$ and $h_3$, we obtain the perturbation equation for $\phi$
\begin{eqnarray}\label{even3}
&&\left(\frac{\omega^2}{U^2}-\frac{\phi^{'4}f^2}{8f^{'2}}+\frac{3\phi^{'2}}{2}+\frac{f\phi^{'}\phi^{''}}{f^{'}}
+\frac{fU^{'}\phi^{'2}}{2f^{'}U}-\frac{\partial^2_{\phi}V}{U}\right)\psi
\nonumber\\&&+\psi^{''}+\left(\frac{U^{'}}{U}+\frac{2f^{'}}{f}\right)\psi^{'}=0\;.
\end{eqnarray}

Similar to the case of odd parity, by introducing a new dynamical variable $\Psi$ taking the place of $\psi$
\begin{eqnarray}
\Psi\equiv {f}\psi\;,
\end{eqnarray}
and the tortoise coordinate $r_{\ast}$, we find the perturbation equation Eq.~(\ref{even3}) is simplified to be
\begin{eqnarray}
-\partial^2_{r_{\ast}}\Psi\left(r_{\ast}\right)+V_{eff}\left(r_{\ast}\right)\Psi\left(r_{\ast}\right)=\omega^2\Psi\left(r_{\ast}\right)\;,
\end{eqnarray}
where
\begin{eqnarray}
&&V_{eff,even}\left(r_{\ast}\right)=\frac{1}{8}\cdot\frac{\phi^{'4}f^2U^2}{f^{'2}}-\frac{5}{4}U^2\phi^{'2}-\frac{\phi^{'}fU^2\phi^{''}}{f^{'}}
\nonumber\\&&-\frac{\phi^{'2}fUU^{'}}{2f^{'}}+\frac{f^{'}UU^{'}}{f}+U\partial_{\phi}^2V\left(\phi\right)\;,
\end{eqnarray}
and the prime denotes the derivative with respect to $x$.

\begin{figure}[htbp]
	\centering
	\includegraphics[width=8cm,height=6cm]{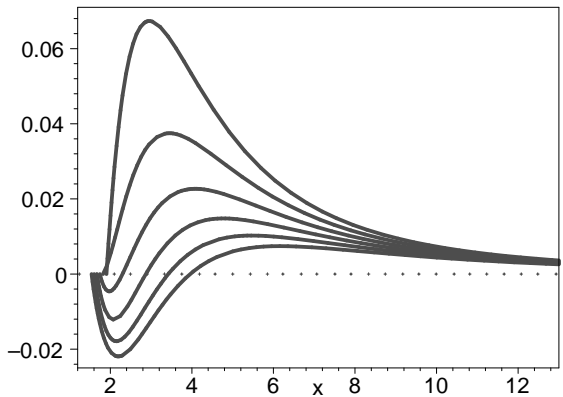}
	\caption{The effective potential $V_{eff,even}$ of monopole perturbation for even parity when $M=1$ and $w_2=0.5,\ 1, \ 1.5, \ 2.0, \ 2.5,\ 3.0$, respectively, from top to down.}
	\label{Vex.eps}
\end{figure}

 In Fig.~(\ref{Vex.eps}), we plot the effective potential $V_{eff,even}$ for the monopole perturbations in terms of $x$. We find that the potential is positive everywhere and tends to zero at both the infinity and the horizon when $w_2\leq 1$ with $M=1$. On the other hand, when $w_2>1$ with $M=1$, the effective potential would develop a negative potential in the vicinity of black hole event horizon. This means there would exist bound state in the vicinity of horizon. The bound states correspond negative eigenvalues of $\omega^2$. Thus an unstable mode is produced. Furthermore, if the mass $M$ vanishes, the effective potential $V_{eff,even}$ is always negative outside the black hole. Therefore, the black holes are always unstable in this case.

 As a conclusion, we find the parameter space of $w_2$ for stability is
\begin{eqnarray}
0\leq w_2<1.1M^2\;.
\end{eqnarray}

\section{scalar perturbations}
In section IV, we studied the stability problem of black holes using the method of metric perturbations and scalar field perturbations. Except for this method, there is the other method, the so-called testing-field method. In this situation, the fields, such as scalar field, Maxwell field and Dirac field, evolving in the background of black hole is regarded as the perturbations to the spacetime while the background spacetime itself is not perturbed. In this section we study the stability of black holes using the testing-field method. For simplicity, we consider a scalar field which obeys the Klein-Gordon equation
\begin{eqnarray}
\nabla^2\Phi+\mu_0^2\Phi=0\;,
\end{eqnarray}
where $\mu_0$ is the rest mass of the scalar particles.
Here $\nabla^2$ is the four dimensional Laplace operator and $\Phi$ the massless scalar field. Making the standard decomposition
\begin{eqnarray}
\Phi=e^{-I\omega t}Y_{lm}\left(\theta,\ \phi\right)\frac{\Psi\left(r\right)}{r}\;,
\end{eqnarray}
we obtain the radial perturbation equation
\begin{eqnarray}
\frac{d^2\Psi}{d r_{\ast}^2}+\left(\omega^2-V_{eff,s}\right)=0\;,
\end{eqnarray}
where the effective potential is given by
\begin{eqnarray}
V_{eff,s}=U\left[\frac{l\left(l+1\right)}{f^2}+\frac{\left(Uf^{'}\right)^{'}}{f}+\mu_0^2\right]\;,
\end{eqnarray}
where the prime denotes the derivative with respect to $x$. We find the effective potential is everywhere positive in the exterior of the black hole. Therefore,
the black hole is stable to scalar perturbations.

\section{black hole thermodynamics}

Finally, we make an investigation on the thermodynamics for the one-scalar-hair black hole. Hawking showed that for the outermost, event horizon in an asymptotically flat spacetime, the
temperature of black hole is
\begin{equation}
T_{EH}=\frac{\kappa_{EH}}{2\pi}\;,
\end{equation}
where the surface gravity $\kappa_{EH}$ is defined by evaluating
\begin{equation}
l^{\mu}\nabla_{\mu}l^{\nu}=\kappa l^{\nu}\;,
\end{equation}
on the event horizon. Here $l^{\mu}$ is the future-directed null generator of the event horizon, which coincides
with a Killing vector $K^{\mu}$ on the horizon. If the Killing vector is adopted as $K^{\mu}=\partial/\partial t$, then we have
\begin{equation}
\kappa_{EH}=\frac{1}{2}\cdot\frac{U_{,r}}{\sigma}|_{r=r_{EH}}\;.
\end{equation}

As a result, the black hole temperature takes the form
\begin{eqnarray}
T_{EH}&=&\frac{1}{4\pi}\cdot\frac{U_{,r}}{\sigma}|_{r=r_{EH}}\;.
\end{eqnarray}
The radius $r_{EH}$ of event horizon is determined by
 \begin{eqnarray}\label{eq:t1}
U\mid_{r=r_{EH}}=0\;.
\end{eqnarray}
Thus the temperature is eventually determined by the mass $M$ and the scalar charge $w_2$.

The entropy of black holes generally satisfies the area law which states that the entropy is a quarter of the area of black hole event horizon \cite{beck:1973,haw:1974,gib:1977}.
Therefore we have the entropy of the black hole
 \begin{equation}
S_{EH}=\pi r_{EH}^2\;.
\end{equation}
Bardeen et. al \cite{bar:1973} have shown that the static black holes satisfy the Smarr formula
\begin{equation}
M=2T_{EH}S_{EH}+M^{{(ext)}}\;,
\end{equation}
where
\begin{equation}
M^{{(ext)}}=-\int_{r\geq r_{EH}}d^3x\sqrt{-g}\left(2T_0^0-T_{\mu}^{\mu}\right)\;,
\end{equation}
is the contribution of the scalar field outside the event horizon to the total mass $M$. In order to check this formula, we should compute the exterior mass $M^{{(ext)}}$. To this end, we resort to the Einstein equations
\begin{equation}
G_{\mu\nu}=8\pi T_{\mu\nu}\;.
\end{equation}
Substituting the metric functions Eq.~(\ref{41}) and Eq.~(\ref{42}) into the Einstein equations, we obtain
\begin{eqnarray}
M^{{(ext)}}&=&-\int_{r\geq r_{EH}}4\pi r^2\sigma\left[\frac{1}{8\pi}\left(G_0^0-G_1^1-2G_2^2\right)\right]dr\nonumber\\
&=&-\frac{2w_2\left(5w_2+6Mr_{EH}\right)}{15r_{EH}\left(r_{EH}^2+w_2\right)}<0\;.
\end{eqnarray}
We remember that $w_2$ is positive. So the mass contributed by the scalar field is negative.
Substituting the expressions of $M^{{(ext)}}$, the Hawking temperature $T_{EH}$ and the entropy $S_{EH}$ into the Smarr formula, we find it is indeed satisfied.

In the next, let's consider the first law of black hole thermodynamics. Motivated  by the procedure adopted in the investigations of black hole thermodynamics with non-linear Maxwell field \cite{yu:2020}, we find we should introduce two conjugated thermodynamical variables, $\bar{\alpha}$ and $\mathbf{A}$ with
\begin{eqnarray}
\bar{\alpha}\equiv\frac{w_2^{\frac{3}{4}}\left(5r_{EH}^2+3w_2\right)^{\frac{1}{2}}}{r_{EH}^{\frac{1}{4}}\left(3r_{EH}^2+2w_2\right)^{\frac{5}{8}}}\;,
\end{eqnarray}
\begin{eqnarray}
\mathbf{A}\equiv-\frac{2}{3}\cdot\frac{w_2^{\frac{1}{4}}\left(3r_{EH}^2+2w_2\right)^{\frac{13}{8}}}{r_{EH}^{\frac{3}{4}}\left(5r_{EH}^2+3w_2\right)^{\frac{3}{2}}}\;,
\end{eqnarray}
where $\bar{\alpha}$ has the dimension of energy and $\mathbf{A}$ is dimensionless.

Then the Smarr formula takes the form

\begin{equation}
M=2T_{EH}S_{EH}+\mathbf{A}\bar{\alpha}\;,
\end{equation}
namely, the exterior mass is expressed as
\begin{equation}
M^{(ext)}=\mathbf{A}\bar{\alpha}\;.
\end{equation}
What is the physical significance of $\mathbf{A}$ and $\bar{\alpha}$? Considering the case of black hole solution with vanishing mass, one has the event horizon $r_{EH}\sim \sqrt{w_2}$. So we have $\bar{\alpha}\sim\sqrt{w_2}$ and $\mathbf{A}$ a pure real number. We conclude $\bar{\alpha}$ is associated to the scalar hair of the black hole and $\mathbf{A}$ is a dimensionless constant.

Bearing in mind the  total mass$M$, the Hawking temperature $T_{EH}$, the Bekenstein-Hawking entropy $S_{EH}$, the dimensionless constant $\mathbf{A}$ and the scalar hair $\bar{\alpha}$ are the functions of event horizon $r_{EH}$ and $w_2$, we obtain the first law of thermodynamics
\begin{equation}
dM=T_{EH}dS_{EH}+\mathbf{A}d\bar{\alpha}\;.
\end{equation}
\section{Conclusion and discussion}
In conclusion, starting from the most general expression, namely, an infinite series of $\phi$ for the scalar potential and by using the Taylor
series method, we find the static and asymptotically flat black hole solutions are eventually determined by two integration constants and many coupling constants. One of the integration constants is the physical mass and the other can be understood as the scalar charge. The coupling constants originate from the scalar potential.
Different from the mass, there is no symmetry correspondence for the scalar charge and the coupling constants. Therefore, they are not conserved and one cannot understand them as Noether charges. But since both the scalar charge and the coupling constants are present in the metric functions, they must play great role in the structure of black hole spacetime. Then it is reasonable to regard them black hole scalar hairs. So a black hole can have many hairs. By using the solution-generating method, we arrive at above conclusions once again. This shows the two methods are consistent with each other.

The no-scalar-hair theorem of Bekenstein \cite{bek:1972l,bek:1972da,bek:1972db} states that the static black holes with the energy momentum tensor contributed by a minimally coupled canonical scalar field and a scalar potential can not have the scalar hairs. Abandoning the third assumption made in the theorem is the price we pay in order to obtain scalar hairy black holes. The third assumption requires the potential to be everywhere non-negative. The well-known Higgs potential tells us this may be not necessary. We find the scalar potential becomes negative merely for sufficient large $\phi$ or in the interior of black hole. On the other hand, the black holes are stable to both odd parity perturbations and scalar perturbations. As for even parity perturbations, there still remains parameter space for the stability of the black holes. Therefore, these scalar hairy black holes are physically significant.

Finally, we make an investigation on the thermodynamics of scalar hairy black holes. Both the Smarr formula and the first law of thermodynamics are developed. We find that a new pair of conjugated thermodynamical variables should be defined except for the pair of Hawking temperature and Bekenstein-Hawking entropy. One of the new variables is determined by the physical mass and scalar charge and the other one is dimensionless.

\section*{ACKNOWLEDGMENTS}

This work is partially supported by the Strategic Priority Research Program ``Multi-wavelength Gravitational Wave Universe'' of the
CAS, Grant No. XDB23040100 and the NSFC under grants 11633004, 11773031.

\newcommand\arctanh[3]{~arctanh.{\bf ~#1}, #2~ (#3)}
\newcommand\ARNPS[3]{~Ann. Rev. Nucl. Part. Sci.{\bf ~#1}, #2~ (#3)}
\newcommand\AL[3]{~Astron. Lett.{\bf ~#1}, #2~ (#3)}
\newcommand\AP[3]{~Astropart. Phys.{\bf ~#1}, #2~ (#3)}
\newcommand\AJ[3]{~Astron. J.{\bf ~#1}, #2~(#3)}
\newcommand\GC[3]{~Grav. Cosmol.{\bf ~#1}, #2~(#3)}
\newcommand\APJ[3]{~Astrophys. J.{\bf ~#1}, #2~ (#3)}
\newcommand\APJL[3]{~Astrophys. J. Lett. {\bf ~#1}, L#2~(#3)}
\newcommand\APJS[3]{~Astrophys. J. Suppl. Ser.{\bf ~#1}, #2~(#3)}
\newcommand\JHEP[3]{~JHEP.{\bf ~#1}, #2~(#3)}
\newcommand\JMP[3]{~J. Math. Phys. {\bf ~#1}, #2~(#3)}
\newcommand\JCAP[3]{~JCAP {\bf ~#1}, #2~ (#3)}
\newcommand\LRR[3]{~Living Rev. Relativity. {\bf ~#1}, #2~ (#3)}
\newcommand\MNRAS[3]{~Mon. Not. R. Astron. Soc.{\bf ~#1}, #2~(#3)}
\newcommand\MNRASL[3]{~Mon. Not. R. Astron. Soc.{\bf ~#1}, L#2~(#3)}
\newcommand\NPB[3]{~Nucl. Phys. B{\bf ~#1}, #2~(#3)}
\newcommand\CMP[3]{~Comm. Math. Phys.{\bf ~#1}, #2~(#3)}
\newcommand\CQG[3]{~Class. Quantum Grav.{\bf ~#1}, #2~(#3)}
\newcommand\PLB[3]{~Phys. Lett. B{\bf ~#1}, #2~(#3)}
\newcommand\PRL[3]{~Phys. Rev. Lett.{\bf ~#1}, #2~(#3)}
\newcommand\PR[3]{~Phys. Rep.{\bf ~#1}, #2~(#3)}
\newcommand\PRd[3]{~Phys. Rev.{\bf ~#1}, #2~(#3)}
\newcommand\PRD[3]{~Phys. Rev. D{\bf ~#1}, #2~(#3)}
\newcommand\RMP[3]{~Rev. Mod. Phys.{\bf ~#1}, #2~(#3)}
\newcommand\SJNP[3]{~Sov. J. Nucl. Phys.{\bf ~#1}, #2~(#3)}
\newcommand\ZPC[3]{~Z. Phys. C{\bf ~#1}, #2~(#3)}
\newcommand\IJGMP[3]{~Int. J. Geom. Meth. Mod. Phys.{\bf ~#1}, #2~(#3)}
\newcommand\IJTP[3]{~Int. J. Theo. Phys.{\bf ~#1}, #2~(#3)}
\newcommand\IJMPD[3]{~Int. J. Mod. Phys. D{\bf ~#1}, #2~(#3)}
\newcommand\IJMPA[3]{~Int. J. Mod. Phys. A{\bf ~#1}, #2~(#3)}
\newcommand\GRG[3]{~Gen. Rel. Grav.{\bf ~#1}, #2~(#3)}
\newcommand\EPJC[3]{~Eur. Phys. J. C{\bf ~#1}, #2~(#3)}
\newcommand\PRSLA[3]{~Proc. Roy. Soc. Lond. A {\bf ~#1}, #2~(#3)}
\newcommand\AHEP[3]{~Adv. High Energy Phys.{\bf ~#1}, #2~(#3)}
\newcommand\Pramana[3]{~Pramana.{\bf ~#1}, #2~(#3)}
\newcommand\PTEP[3]{~PTEP.{\bf ~#1}, #2~(#3)}
\newcommand\PTP[3]{~Prog. Theor. Phys{\bf ~#1}, #2~(#3)}
\newcommand\APPS[3]{~Acta Phys. Polon. Supp.{\bf ~#1}, #2~(#3)}
\newcommand\ANP[3]{~Annals Phys.{\bf ~#1}, #2~(#3)}
\newcommand\RPP[3]{~Rept. Prog. Phys. {\bf ~#1}, #2~(#3)}
\newcommand\ZP[3]{~Z. Phys. {\bf ~#1}, #2~(#3)}
\newcommand\NCBS[3]{~Nuovo Cimento B Serie {\bf ~#1}, #2~(#3)}
\newcommand\AAP[3]{~Astron. Astrophys.{\bf ~#1}, #2~(#3)}
\newcommand\MPLA[3]{~Mod. Phys. Lett. A.{\bf ~#1}, #2~(#3)}
\newcommand\NT[3]{~Nature.{\bf ~#1}, #2~(#3)}
\newcommand\PT[3]{~Phys. Today. {\bf ~#1}, #2~ (#3)}
\newcommand\APPB[3]{~Acta Phys. Polon. B{\bf ~#1}, #2~(#3)}
\newcommand\NP[3]{~Nucl. Phys. {\bf ~#1}, #2~ (#3)}
\newcommand\JETP[3]{~JETP Lett. {\bf ~#1}, #2~(#3)}

\end{document}